\documentclass[10pt,a4paper]{article}
\usepackage{jheppub}
\usepackage{amsmath,amssymb}
\usepackage{bm}
\usepackage{tensor}
\usepackage{physics,microtype}
\usepackage[dvipsnames]{xcolor}

\usepackage{orcidlink}    
\numberwithin{equation}{section}

\renewcommand\thesection{\arabic{section}}

\begin{document}

\title{Supergravity with Lagrange Multiplier Fields in $2+1$ Dimensions}

\author[a,b]{D.~G.~C.~McKeon\orcidlink{0000-0002-6152-4495},}
\emailAdd{dgmckeon2@uwo.ca}

\author[c]{F.~T.~Brandt\orcidlink{0000-0001-7873-7684},}
\emailAdd{fbrandt@usp.br}

\author[c]{J.~Frenkel\orcidlink{0000-0003-1178-1001},}
\emailAdd{jfrenkel@if.usp.br}

\author[c,1]{and S.~Martins-Filho\orcidlink{0000-0003-4195-2713} %
\note{Correspoding author. Now at Instituto de F\'{\i}sica Te\'orica, Universidade Estadual Paulista (UNESP), Rua Dr. Bento Teobaldo Ferraz, 271 - Bloco II, 01140-070 S\~ao Paulo, SP, Brazil.}}
\emailAdd{s.martins-filho@unesp.br}

\affiliation[a]{Department of Applied Mathematics, The University of Western Ontario, London, Ontario N6A 5B7, Canada}
\affiliation[b]{Department of Mathematics and Computer Science, Algoma University, Sault Ste.\ Marie, Ontario P6A 2G4, Canada}
\affiliation[c]{Instituto de F\'{\i}sica, Universidade de S\~ao Paulo, S\~ao Paulo, SP 05508-090, Brazil}

\abstract{
We examine the first-order Einstein--Cartan (EC) action in $2+1$ dimensions, including a cosmological term and its supersymmetric extension. In this setting the spin connection can be expressed as an axial vector, yielding an action that is bilinear in the quantum fields and allows quantization without background fields. From the complete set of first-class constraints, the associated gauge transformations all follow. These differ from the standard supersymmetric diffeomorphism and local Lorentz invariances. Using the closed gauge algebra, we construct the Faddeev--Popov--Nielsen path integral and then show how a Lagrange multiplier field can be introduced to remove higher-loop contributions while preserving unitarity and gauge invariance. This indicates the possibility of being able to use a Lagrange multiplier field to formulate a unitary and renormalizable model for Einstein–Cartan gravity coupled to the Standard Model in $3+1$ dimensions.}


\arxivnumber{2601.10593}

\maketitle

\section{Introduction}

Attempts to construct a unified, renormalizable and unitary theory of the
fundamental interactions often require going beyond the Einstein--Hilbert
description of gravity. The Einstein--Hilbert action coupled to bosonic matter
is not renormalizable, but it has been shown that by introducing suitable
Lagrange multiplier fields one can enforce the equations of motion 
at the classical level. At the quantum level, this results in
a renormalizable and unitary theory that reduces to classical general
relativity when quantum corrections are neglected.
These quantum corrections correspond to the usual one-loop effects that arise when
the Lagrange multiplier (LM) fields are absent.

Including matter with spin requires replacing the Einstein--Hilbert action by
the Einstein--Cartan (EC) action, whose canonical and gauge structure in
$3+1$ dimensions remains difficult to disentangle. In contrast, the
$2+1$-dimensional model is far simpler: the spin connection can be dualized to
an axial vector, the action contains a quadratic term in the quantum fields, and
the first-class constraints can be identified explicitly. These features make
$2+1$ dimensions an ideal setting in which to analyse the gauge structure and
quantization of the EC model, as well as the role of a Lagrange multiplier in
eliminating higher-loop contributions.
This use of a lower dimensional model of gravity to gain insight into gravity in $3+1$D
has long been standard practice \cite{Witten1988,ref23}. Supergravity models in 2+1D have been extensively studied in order to gain insight into supergravity in 3+1D (\cite{Alkac2015}, for example).

The first-order (``Palatini'') form of the
Einstein--Cartan (EC) action in $3+1$ dimensions is
\cite{Cartan1922,Kibble1961,Hehl1976}
\begin{equation}
S_{\text{EC}4} = -\frac{1}{\kappa^{2}} \int d^{4}x \; e  R(e,\omega)
\label{eq:1}
\end{equation}
where $\kappa^{2} = 16\pi G$ ($G$ = Newton's constant),
\begin{equation}
e = \det e_{\mu}^{\; a} = \bigl( - \det g_{\mu\nu} \bigr)^{1/2} \quad
(e_{\mu}^{\; a} = \text{``vierbein''}; 
 g_{\mu\nu} = \eta_{ab}  e_{\mu}^{\; a} e_{\nu}^{\; b} \mbox{ with }
 \eta_{ab} = \mathrm{diag}(+---)),
\label{eq:2}
\end{equation}
\begin{equation}
R_{\mu\nu ab}
= \partial_{\mu} \omega_{\nu ab}
- \partial_{\nu} \omega_{\mu ab}
+ \omega_{\mu ac}  \omega_{\nu}{}^{c}_{\;b}
- \omega_{\nu ac}  \omega_{\mu}{}^{c}_{\;b},
\label{eq:3}
\end{equation}
where $\omega_{\mu\; ab} = - \omega_{\mu\; ba}$ is the ``spin connection''
and
\begin{equation}
R(e,\omega) = e^{a\mu}  e^{b\nu}  R_{\mu\nu ab} \; .
\label{eq:4}
\end{equation}
One can show that the dynamics that follows from $S_{\text{EC}4}$ also
follows from 
\begin{equation}
S^\prime_{\text{EC}4} = -\frac{1}{L^2}
\int d^{4}x \; \varepsilon^{\mu\nu\lambda\sigma} \varepsilon_{abcd}\;
e^{a}_{\; \mu} e^{b}_{\; \nu} R_{\lambda\sigma}^{\;\;\; cd}  \;  .
\label{eq:5}
\end{equation}
We scale the dimensionful parameter $L$ to equal one.

It is immediately apparent that the action $S_{\text{EC}4}$ is invariant
under two local gauge invariances. The first is local
infinitesimal diffeomorphism invariance (DI), in which
\begin{equation}\label{eq:6}
x^{\mu} \to x^{\mu} + \xi^{\mu}(x)
\end{equation}
so that a space–time vector $V_{\mu}$ transforms as
\begin{subequations}\label{eq:7}
\begin{equation}
\delta V_{\mu} = - \xi^{\rho} \partial_{\rho} V_{\mu}
- V_{\rho} \partial_{\mu} \xi^{\rho}
\label{eq:7a}
\end{equation}
and
\begin{equation}
\delta V^{\mu} = - \xi^{\rho} \partial_{\rho} V^{\mu}
                 + V^{\rho} \partial_{\rho} \xi^{\mu}.
\label{eq:7b}
\end{equation}
\end{subequations}
The second transformation is local Lorentz invariance (LLI), in which a
tangent–space vector $A_{a}$ transforms as
\begin{equation}
A_{a} \rightarrow A_{a} + \lambda_{a}{}^{b} A_{b} 
\quad (\lambda_{ab}(x) = - \lambda_{ba}(x)).
\label{eq:8}
\end{equation}
The transformations of Eqs.~\eqref{eq:7} and \eqref{eq:8} can be used to discuss
the quantization \cite{Brandt2024a} and renormalization \cite{Brandt2024b}
of $S_{\text{EC}4}$.

However, the non-linear nature of $S_{\text{EC}4}$ presents two
immediate difficulties. The first is that it is necessary to use
background field quantization \cite{DeWitt1967,Abbott1981} to compute quantum
corrections to the classical action, as only in the presence of
a background field can a contribution to $S_{\text{EC}4}$ that is
quadratic in the quantum field be obtained.

A second problem is that the invariances of Eqs.~\eqref{eq:7} and \eqref{eq:8},
though manifest, are not necessarily related to the
canonical structure of the theory. Dirac \cite{Dirac1950,DiracQM} established
a connection between the first class constraints and the
presence of local gauge invariances. Two ways of deriving
the explicit form of these local gauge invariances from
the primary first class constraints are provided in
refs.~\cite{Castellani1982,Henneaux1990}. Applying either of these procedures
to $S_{\text{EC}4}$ in Eq.~\eqref{eq:1} is very difficult due to the complexity of
identifying the primary first class constraints in the theory,
but there are indications \cite{Kiriushcheva:2010tgs,Leclerc2007,McKeon2010}
that these constraints do not lead to the DI of Eq.~\eqref{eq:7a},
but rather to a form of translational invariance (TI) in the tangent space.

We are thus motivated to examine more closely the EC
action in $2+1$ dimensions. We will extend our considerations
to also incorporate a cosmological term as well as
a supersymmetric interaction with a ``gravitino'' spinor field
\cite{Witten1988,HoweTucker1978,AchucarroTownsend1986,MatschullNicolai1994,Hayashi1992}. By going to one
lower spatial dimension, we
find that the action has a term bilinear in the
quantum fields, making it possible to avoid the necessity of
introducing background fields. It also provides a way of
deriving local gauge invariances from the primary first-class
constraints that arise \cite{McKeon2014,McKeon2020,Frolov2012}.

The simplification that occurs when one goes from
3+1 to 2+1 dimensions arises principally because the
antisymmetric spin connection $\omega_{\mu ab} = - \omega_{\mu ba}$ can be
replaced by an axial vector $\omega_{\mu i} = \varepsilon_{ijk} \omega_{\mu}^{\; jk}$.
The action is based on $S'_{\text{EC}4}$ of Eq.~\eqref{eq:5}
\cite{McKeon2020}
\begin{equation}
\label{eq:9}\begin{split}
S_{\text{EC}3}
&= \int d^{3}x  \varepsilon^{\mu\nu\rho} \Bigl[
b^{i}_{\mu} R_{\nu\rho i}
+ \bar{\psi}_{\mu} D_{\nu} \psi_{\rho}
+ \frac{\Lambda}{3} \varepsilon_{ijk} b^{i}_{\mu} b^{j}_{\nu} b^{k}_{\rho}
+ \frac{i\varkappa}{2} \bar{\psi}_{\mu} b_{\nu}^{\; i} \gamma_i \psi_{\rho}
\Bigr] \\&\equiv \int d^{3}x  \mathcal{L}_{\text{EC}3}
\end{split}
\end{equation}
with $\Lambda$ being a ``cosmological constant'' and $\varkappa$ is a dimensionless parameter (see
Appendix \ref{appA} for notation). Interesting black hole solutions occur in $2+1$ dimensional space-times in which $\Lambda$ is a negative constant \cite{ref23}.

The structure of the paper is as follows. In Sec.~2 we examine the gauge
invariances that arise from the first-class constraints of the
Einstein--Cartan action in $2+1$ dimensions, including the corresponding
transformations and the closure of the gauge algebra.
This gauge algebra has some novel features, such as having the composition of two
Fermionic gauge transformations leading to a Bosonic gauge transformation (Eq. \eqref{eq:13} below).
In Sec.~3 we turn to the
relation between the first- and second-order formulations of the theory,
showing how the spin connection can be eliminated in favour of its shifted
counterpart and how this affects the form of the action.
A novel feature of this second order form is that a rotation in the target space
is a local symmetry, but it is not associated with a gauge transformation.
Sec.~4 presents the
quantization of the model using the Faddeev--Popov--Nielsen procedure and the
derivation of the associated BRST transformations.
These transformations lead to a new way of analyzing the dependence of Green's
functions on the gauge parameters.
In Sec.~5 we introduce a
Lagrange multiplier field and show how it can be used to restrict radiative
corrections to one-loop order while preserving gauge invariance and unitarity.
Finally, Sec.~6 contains our concluding remarks and discusses the prospects of
extending these results to the $3+1$-dimensional Einstein--Cartan theory.

\section{Gauge Invariance}
In refs.~\cite{McKeon2014,McKeon2020} it is shown that the first
class constraints that follow from $S_{\text{EC}3}$ result in the local
gauge invariances
\begin{subequations}\label{eq:10}\allowdisplaybreaks
\begin{equation}
\delta b^{i}_{\mu}
= {\cal D}_{\mu}^{ij} A_{j} - \varepsilon^{ijk} b_{\mu j} B_{k}
+ \frac{i}{2} \bar{C}\gamma^{i}\psi_{\mu},
\label{eq:10a}
\end{equation}
\begin{equation}
\delta \omega^{i}_{\mu}
= {\cal D}_{\mu}^{ij} B_{j}
- \varkappa^2 \varepsilon^{ijk} b_{\mu j} A^{k}
+ \frac{i \varkappa}{2} \bar{C}\gamma^{i}\psi_{\mu},
\label{eq:10b}
\end{equation}
\begin{equation}
\delta \psi_{\mu}
= D_{\mu} C
+ i\frac{\varkappa}{2} \gamma^{j} b_{\mu j} C
- \frac{i}{2} \gamma^j(\varkappa A_{j}+B_{j})\psi_{\mu}.
\label{eq:10c}
\end{equation}
\end{subequations}
These three transformations all show that the longitudinal components of 
$b_{\mu i}$, $\omega_{\mu i}$, and $\psi_{\mu}$ are gauge artifacts.

The gauge functions $A_{i}$ and $B_{i}$ are bosonic while $C$ is a
Grassmann Majorana spinor. The gauge transformations
of Eq.~\eqref{eq:10} close if
\begin{equation}
\Lambda = -\varkappa^{2}.
\label{eq:11}
\end{equation}
If a gauge transformation $\delta_{i}$ is associated with gauge
functions $A_{i}, B_{i}, C_{i}$, then from Eq.~\eqref{eq:10} it follows that
\begin{equation}
(\delta_{1}\delta_{2} - \delta_{2}\delta_{1}) = \delta_{T},
\label{eq:12}
\end{equation}
where
\begin{subequations}\label{eq:13}\allowdisplaybreaks
\begin{equation}
A_{T_i}
= \varepsilon_{ijk}\Bigl( B^{j}_{1} A^{k}_{2}+ A^{j}_{1} B^{k}_{2} \Bigr)
+ \frac{i}{2}\bar{C}_{1}\gamma_{i} C_{2},
\label{eq:13a}
\end{equation}
\begin{equation}
B_{T_i}
= \varepsilon_{ijk}
\Bigl(\varkappa^2 A_{1}^{j} A_{2}^{k} + B^{j}_{1} B^{k}_{2}\Bigr)
+ \frac{i\varkappa}{2} \bar{C}_{1} \gamma_{i} C_{2},
\label{eq:13b}
\end{equation}
\begin{equation}
C_{T}
= \frac{i\varkappa}{2}
\Bigl(
A^{j}_{2} \gamma_{j} C_{1}- A^{j}_{1} \gamma_{j} C_{2}
\Bigr)
+ \frac{i}{2}
\Bigl(B^{j}_{2} \gamma_{j} C_{1}- B^{j}_{1} \gamma_{j} C_{2}\Bigr).
\label{eq:13c}
\end{equation}
\end{subequations}
From Eqs. \eqref{eq:13a} and \eqref{eq:13b} 
we see that commuting two Fermionic gauge transformations
leads to a Bosonic gauge transformation, which is a feature of supersymmetry.
The invariances of Eq.~\eqref{eq:10} are quite distinct from
the manifest invariances of
Eqs.~\eqref{eq:7} and \eqref{eq:8}.  
It is interesting to derive two invariances of
$S_{\text{EC}3}$ by considering what follows from Eq.~\eqref{eq:7} if we  
consider
\begin{subequations}\label{eq:14}\allowdisplaybreaks
\begin{equation}
A_{i} = - b_{\mu i} \xi^{\mu},
\label{eq:14a}
\end{equation}
\begin{equation}
B_{i} = - \omega_{\mu i} \xi^{\mu},
\label{eq:14b}
\end{equation}
\begin{equation}
C = - \xi^{\mu} \psi_{\mu}.
\label{eq:14c}
\end{equation}
\end{subequations}
as in Refs. \cite{Kiriushcheva:2010tgs,Leclerc2007}. We find that Eq.~\eqref{eq:10} becomes
\begin{subequations}\label{eq:15}
\begin{equation}
\delta b^{i}_{\mu}
= - \bigl[
b^{i}_{\lambda} \partial_{\mu} \xi^{\lambda}
+ \xi^{\lambda} \partial_{\lambda} b^{\;\; i}_{\mu}
\bigr]
+  \xi^\lambda \bigl[
  \partial_\lambda b_\mu^{\;\; i} - \partial_\mu b_{\lambda}^{\;\; i} +\varepsilon^{ijk}(
\omega_{\mu j} b_{\lambda k} - b_{\mu j} \omega_{\lambda k})- \frac{i}{2} \bar\psi_\lambda \gamma^i\psi_\mu 
  \bigr],
\label{eq:15a}
\end{equation}
\begin{equation}
\delta \omega^{i}_{\mu}
= - \bigl[
\omega^{i}_{\lambda} \partial_{\mu} \xi^{\lambda}
+ \xi^{\lambda} \partial_{\lambda} \omega^{i}_{\mu}
\bigr]
+\xi^\lambda\bigl[
R_{\lambda\mu}{}^{\; i}+\varkappa^2\varepsilon^{ijk} b_{\mu j} b_{\lambda k} 
- \frac{i\varkappa}{2} \bar{\psi}_\lambda \gamma^{i} \psi_{\mu}\bigr] ,
\label{eq:15b}
\end{equation}
\begin{equation}
\delta \psi_{\mu}
= - \bigl[
\psi_{\lambda} \partial_{\mu} \xi^{\lambda}
+ \xi^{\lambda} \partial_{\lambda} \psi_{\mu}
\bigr]
+ \xi^\lambda\bigl[{D}_{\lambda} \psi_\mu
+\frac{i\varkappa}{2} \gamma^j b_{\lambda j} \psi_\mu
-{D}_{\mu} \psi_\lambda
-\frac{i\varkappa}{2} \gamma^j b_{\mu j} \psi_\lambda
\bigr].
\label{eq:15c}
\end{equation}
\end{subequations}

The first two terms in each of the transformations in Eq.~\eqref{eq:15}
correspond to the diffeomorphism transformation of Eq.~\eqref{eq:7a},
with $V_{\mu}$ being identified with $b^{i}_{\mu}$, $\omega^{i}_{\mu}$ and
$\psi_{\mu}$ respectively.
The remaining terms in Eq.~\eqref{eq:15} also correspond therefore to
transformations that leave $S_{\text{EC}3}$ invariant; they might be
considered as being ``translations'' in the tangent space. Alone,
they do not follow directly from the first class constraints present
in this model.

\section{The Second Order Form}
In order to show how the first order (Palatini) form
of $S_{\text{EC}3}$ in Eq.~\eqref{eq:9} is related to the second order form, we
use the approach of ref.~\cite{Brandt:2025lkd}, where the Einstein--Hilbert
(EH) action is considered. First, we note that the
equation of motion for $\omega^{i}_\lambda$ that follows from $S_{\text{EC}3}$ is
\begin{equation}
\varepsilon^{\mu\nu\lambda}
\left[
\frac12 \bigl( D^{ij}_{\mu} b_{\nu j}
               - D^{ij}_{\nu} b_{\mu j} \bigr)
- \frac i 4 \bar{\psi}_{\mu} \gamma^{i} \psi_{\nu}
\right] = 0
\label{eq:16}
\end{equation}
Upon contracting this equation with $\varepsilon_{ijk}$, we find that
\begin{equation}
\varepsilon_{ijk}
\left[
\frac12 \bigl( \partial_{\mu} b^{i}_{\nu}
               -\partial_{\nu} b^{i}_{\mu} \bigr)
- \frac i 4 \bar{\psi}_{\mu} \gamma^{i} \psi_{\nu}
\right] 
- \frac12
\bigl( \omega_{\mu j} b_{\nu k}
     - \omega_{\nu j} b_{\mu k} \bigr)
= 0
\label{eq:17}
\end{equation}
Using Eq. \eqref{eq:A5}, Eq.~\eqref{eq:17} leads to
\begin{equation}
\label{eq:18}\begin{split}
\omega_{\nu k}&\equiv \Omega_{\nu k}{}
\\ &= 2\varepsilon_{ijk} b^{\mu j}
\Bigl[
\frac12(\partial_{\mu} b_{\nu}^{\; i} -  \partial_\nu b_{\mu}^{\; i})
- \frac{i}{4}\bar\psi_{\mu}\gamma^{i}\psi_{\nu}
\Bigr]
- b_{\nu k}\!
\left\{
\frac12\varepsilon_{rst} b^{\mu s} b^{\sigma t}
\Bigl[
\frac12(\partial_{\mu} b_{\sigma}^{\; r} - \partial_\sigma b_{\mu}^{\; r})
- \frac{i}{4}\bar\psi_{\mu}\gamma^{r}\psi_{\sigma}
\Bigr]
\right\}.
\end{split}
\end{equation}

A second order form for $S_{\text{EC}3}$ is obtained
by substituting $\Omega^{\mu}_{i}$ for $\omega^{\mu}_{i}$ in Eq.~\eqref{eq:9}.
The resulting action is now second order, but more complicated:
it becomes quadratic in the spinor $\psi_{\mu}$, and the kinetic term
for $b_{\mu i}$ requires expanding $b_{\mu i}$ around a background field.
The shift
\begin{equation}
\omega_{\mu i}^{\prime}
= \omega_{\mu i} - \Omega_{\mu i}
\label{eq:19a}
\end{equation}
in $S_{\text{EC}3}$ removes cross terms in $b_{\mu i}$ and
$\omega_{\mu i}^{\prime}$, as well as terms involving derivatives of
$\omega_{\mu i}$. The kinetic term for $b_{\mu i}$ is obtained only
after expanding $b_{\mu i}$ about a background field.
This term will involve the second derivative of $b_{\mu i}$
From Eq.~\eqref{eq:18}, it follows that $\omega_{\mu i}$ and
$\Omega_{\mu i}$ transform in the same way under the gauge
transformation of Eq.~\eqref{eq:10}, and hence by Eq.~\eqref{eq:10b} we have simply
\begin{equation}
\delta\omega_{\mu}^{\prime}
= -\varepsilon^{ijk} \omega_{\mu j}^{\prime} B_{k}.
\label{eq:20a}
\end{equation}
Thus, under a gauge transformation, the longitudinal part of
$\omega_{\mu}^{\prime }$ is unchanged. Hence
$\omega_{\mu}^{\prime }$
has the unusual situation in which
$\omega_{\mu}^{\prime }$
is a non-dynamical auxiliary field that undergoes a local gauge
transformation that corresponds to a local rotation.
\section{Quantization}
We can quantize using the standard Faddeev–Popov quantization 
procedure \cite{DeWitt1967,FaddeevPopov1967} as by Eq. \eqref{eq:12} the gauge algebra 
is closed. If we work in the ``Lorenz gauge'' (i.e. $ \partial^\mu b_{\mu i} = \partial^\mu \omega_{\mu i} = \partial ^\mu \psi_\mu=0$), then this involves insertion of the constraint factor
\begin{equation}\begin{split}
  I_{\rm FP} = & \int {\cal D}A^i{\cal D}B^i{\cal D}C
{\cal D}p^{\prime i}{\cal D}q^{\prime i}{\cal D}\pi^\prime \\ & \times 
\delta \left\{\partial^\mu\left[\left(
\begin{array}{c}
b_{\mu}{}^{i} \\
\omega_{\mu}{}^{i} \\
\psi_{\mu}
\end{array}\right)\right.\right.
\left.\left.
 +
\left(
\begin{array}{ccc}
{\cal D}^{ij}_\mu & -\varepsilon^{i p j} b_{\mu p} & -\frac{i}{2} \bar{\psi}_\mu\gamma^i \\[4pt]
-\varkappa^{2}\varepsilon^{ipj} b_{\mu p} & {\cal D}^{ij}_\mu & -\frac{i\varkappa}{2} \bar{\psi}_\mu\gamma^i \\[4pt]
-\frac{i\varkappa}{2}\gamma^j\psi_\mu & -\frac{i\gamma^j}{2}{\psi}_\mu  & D_{\mu} + \frac{i \varkappa}{2}\gamma^{k}b_{\mu k}
\end{array}
\right)
\left(
\begin{array}{c}
A_j \\[4pt]
B_j \\[4pt]
C
\end{array}
\right)\right]
-
\left(
\begin{array}{c}
p^{\prime i} \\[4pt]
q^{\prime i} \\[4pt]
\pi^{\prime}
\end{array}
\right) \right\}
\\ & \times
({\det} \; \underset{\sim{}}{M})\!
\int {\cal D} p^i{\cal D} q^i{\cal D}\pi\;
\exp\!\left[i\!\int d^{3}x \left(
\frac{\alpha_p}{2}p^i p_i
+ \frac{\alpha_q}{2}q^i q_i
+ \frac{\alpha_{\pi}}{2}\bar\pi \pi
-p^i p^{\prime}_{ i}-q^i q^{\prime}_{ i}-\bar{\pi} \pi^{\prime}
\right)\right]
\end{split}
\label{eq:19b}
\end{equation}
into the path integral for the generating functional,
where $p^i$, $q^i$ and $\pi$ are Nakanishi–Lautrup (NL) fields, with
$\pi$ being a Grassmann Majorana spinor, and $\underset{\sim{}}{M}$ is the supermatrix
appearing in the argument of the $\delta$-function in Eq. \eqref{eq:19b}.
The resulting generating functional is 
\begin{equation}\begin{split}
&Z[j^{\mu}, J^{\mu i}, K^{\mu}]
=
\int ({\cal D}b_{\mu i}{\cal D}\omega_{\mu i}{\cal D}\psi_{\mu})
({\cal D}p^{i}{\cal D}q^{i}{\cal D}\pi)
({\cal D}c^i{\cal D}d^i{\cal D}e^i{\cal D}f^i)
({\cal D}\phi{\cal D}\chi)
\\
&\times
\exp\!\Biggl\{
i \int d^{3}x \Biggl[
{\cal L}_{\mathrm{EC}3}
+ \frac{\alpha_p}{2}p^{i} p_{i}
+ \frac{\alpha_{q}}{2}q^{i}q_{i}
+ \frac{\alpha_{\pi}}{2}\bar{\pi}\pi
-p^{i}\partial^{\mu}b_{\mu i}
- q^{i}\partial^{\mu}\omega_{\mu i}
-\bar{\pi}\partial^{\mu}\psi_{\mu}
\\ &
+ \begin{pmatrix} e_{i} &  f_{i}&\bar\phi \end{pmatrix}
\partial^{\mu}
\begin{pmatrix}
{\cal D}^{ij}_{\mu} 
&
-\varepsilon^{ipj} b_{\mu p}
&
-\frac{i}{2}\bar{\psi}_{\mu}\gamma^{i}
\\
-\varkappa^{2}\varepsilon^{ipj} b_{\mu p}
&
{\cal D}^{ij}_{\mu}
&
-\frac{i\varkappa}{2}\bar{\psi}_{\mu}\gamma^{i}
\\
-\frac{i\varkappa}{2}\gamma^{j}\psi_{\mu}
&
-\frac{i}{2}\gamma^{j}{\psi}_{\mu}
&
D_{\mu}+\frac{i\varkappa}{2}\gamma^{k}b_{\mu k}
\end{pmatrix}
\begin{pmatrix}
c_j \\ d_j \\ \chi
\end{pmatrix}
\\ 
&+ j^{\mu i} b_{\mu i}
+ J^{\mu i}\omega_{\mu i}
+ \bar{K}^{\mu}\psi_{\mu}
\Biggr]
\Biggr\}.
\end{split}
\label{eq:20b}
\end{equation}
We have Grassmann ghosts \(c_i,d_i,e_i,f_i\), and bosonic ghosts \(\phi,\chi\).
Source fields are \(j^{\mu i}\), \(J^{\mu i}\) and \(K^{\mu}\).

The usual BRST transformations \cite{Becchi1975,Tyutin1975} that follow
from Eqs. \eqref{eq:10} and \eqref{eq:20b} are (with \(\eta\) a Grassmann constant)
\begin{equation}
\delta
\begin{pmatrix}
b_{\mu}{}^{i} \\[4pt]
\omega_{\mu}{}^{i} \\[4pt]
\psi_{\mu}
\end{pmatrix}
=
M^{ij}
\begin{pmatrix}
c_j \\ d_j \\ \chi
\end{pmatrix}
\eta .
\label{eq:21}
\end{equation}
as well as
\begin{equation}
\delta p_{i}
= \delta q_{i}
= \delta\pi
= 0 .
\label{eq:22}
\end{equation}
and
\begin{equation}
\delta e^{ i} = - p^{i}\eta,
\qquad
\delta f^{ i} = - q^{i}\eta,
\qquad
\delta\bar{\phi} = \bar{\pi}{\eta},
\label{eq:23}
\end{equation}

To obtain $\delta c_{i}$, $\delta d_{i}$, $\delta\chi$, we note that for the
argument of the integral in Eq.~\eqref{eq:20b} to remain invariant,
we need to have
\begin{equation}
\delta\!\left[
\underset{\sim{}}{M}
  (b_{\mu i},\omega_{\mu i},\psi_{\mu})
\begin{pmatrix}
c \\ d \\ \chi
\end{pmatrix}
\right]
= 0 .
\label{eq:24}
\end{equation}
We now note that the gauge transformation of Eq.~\eqref{eq:10} is
a realization of
\begin{equation}
\delta_{a}
\begin{pmatrix}
b \\[4pt]
\omega \\[4pt]
\psi
\end{pmatrix}
=
\underset{\sim{}}{M}(b_{\mu i},\omega_{\mu i},\psi_{\mu})
\begin{pmatrix}
A_{l} \\[4pt]
B_{l} \\[4pt]
C
\end{pmatrix},
\label{eq:25}
\end{equation}
It is now apparent that Eqs.~\eqref{eq:12} and \eqref{eq:24} have the same
structure. Upon making use of Eq.~\eqref{eq:21} in Eq.~\eqref{eq:24}, we
find that Eq.~\eqref{eq:24} is satisfied if
\begin{subequations}\label{eq:26}
\begin{align}
\delta c_{i}
&= - \varepsilon^{ijk} d_{j} c_{k}\eta
   + \frac{i}{4}\bar{\chi}\gamma_{i}\chi\eta,
\\[4pt]
\delta d_{i}
&= - \varepsilon^{ijk}
    \left( \frac{\varkappa^{2}}{2}  c_{j} c_{k}
         + \frac 1 2  d_{j} d_{k} \right)\eta
   + \frac{i\varkappa}{4}\bar{\chi}\gamma_{i}\chi\eta,
\\[4pt]
\delta\chi
&= i\frac{\gamma^p}{2}
\left( \varkappa c_p + d_{p} \right)\chi\eta,
\end{align}
\end{subequations}
upon remembering that ($c_{i}, d_{i}$) are fermionic and $\chi$ is bosonic.

The BRST transformations of Eqs.~\eqref{eq:21}, \eqref{eq:22},
\eqref{eq:23}, and \eqref{eq:26}
can be extended to include changes in the gauge parameters
$\alpha_{p}$, $\alpha_{q}$ and $\alpha_{\pi}$ \cite{Nielsen1975}.
This involves introduction of further Grassmann constants
$\rho_{p}, \rho_{q}, \rho_{\pi}$ and the terms
\begin{equation}
\rho_{p} e^{i} p_{i}
\;+\;
\rho_{q} f^{i} q_{i}
\;+\;
\rho_{\pi} \bar{\pi}\chi
\label{eq:27}
\end{equation}
into the effective Lagrangian contained in the argument of the
exponential appearing in
Eq.~\eqref{eq:20b}.  
The BRST transformations are now expanded to include \cite{Nielsen1975}
\begin{equation}
\delta \rho_{p}
=
\delta \rho_{q}
=
\delta \rho_{\pi}
= 0 ,
\label{eq:28}
\end{equation}
and
\begin{equation}
\delta \alpha_{p}
= 2 \rho_{p}\eta,
\qquad
\delta \alpha_{q}
= 2 \rho_{q}\eta,
\qquad
\delta \alpha_{\pi}
= 2 \rho_{\pi}\eta .
\label{eq:29}
\end{equation}
Together, these transformations serve to determine the
dependence of Green's functions on the gauge parameters
$\alpha_{p},\alpha_{q}$ and $\alpha_{\pi}$.
For example, let us see how the Green's function
\begin{equation}
F(x,y)
= \langle 0| T \omega_{\mu i}(x) \omega_{\nu j}(y) |0\rangle ,
\label{eq:30}
\end{equation}
depends on $\alpha_{p}$.  
As the transformations of Eqs.~\eqref{eq:21}, \eqref{eq:22},
\eqref{eq:23}, \eqref{eq:26} and \eqref{eq:29}
leave the theory invariant, we see that
\begin{align}
\biggl[
\delta \alpha_{p}\frac{\partial}{\partial \alpha_{p}}
+ \delta \alpha_{q}\frac{\partial}{\partial \alpha_{q}}
+ \delta \alpha_{\pi}\frac{\partial}{\partial \alpha_{\pi}}
\biggr]
\langle 0| T \omega_{\mu i}(x) \omega_{\nu j}(y) |0\rangle
& \nonumber \\
+ \langle 0| T \delta\omega_{\mu i}(x) \omega_{\nu j}(y) |0\rangle
+ \langle 0| T \omega_{\mu i}(x) \delta\omega_{\nu j}(y) |0\rangle
= 0 .
\label{eq:31}
\end{align}
Upon taking $\partial / \partial \rho_{p}$ of Eq.~\eqref{eq:31}
and then setting $\rho_{p}=\rho_{q}=\rho_{\pi}=0$,
we see that Eqs.~\eqref{eq:27} and \eqref{eq:29} lead to
\begin{align}
&2\eta\frac{\partial}{\partial \alpha_{p}}
  \langle 0|   T \omega_{\mu i}(x) \omega_{\nu j}(y) |0\rangle
  \nonumber \\
&+ \langle 0| T \Big[
\bigl( \partial_{\mu} d_{i}(x)
  - \varepsilon_{ipm}\omega_{\mu}^{p}(x) d^{ m}(x)
   -  \varepsilon_{ipm} b^{p}_\mu(x) c^{m}(x)]
-\frac{i}{2} \bar\psi_\mu(x)\gamma_i \chi(x)\bigr) \eta\Big] \omega_{\nu j}(y)
\Bigl(i\int dz e^k(z) p_k(z)\Bigr)
|0\rangle
 \nonumber \\\nonumber 
&+ \langle 0|T\big[ \omega_{\mu}^{i}(x)\big]
\Big[
\partial_{\nu} d_{j}(y)
- \varepsilon_{jqm}\omega_{\nu}^{ q}(y) d_{\lambda}^{m}(y)
-\varepsilon_{jqm}
b^q_\nu(y) c^m(y)  
  -\frac{i}{2} \bar\psi_\nu(y) \gamma_j \chi(y)\Big] \eta
  \big(i\int dz e^k(z) p_k(z) \big)
|0\rangle \\  &=0.
\label{eq:32}
\end{align}
Eq.~\eqref{eq:32} fixes the dependence of $F(x,y)$ on $\alpha_{p}$.
We note that this method of determining the dependence of a Green's function
on a gauge parameter can be applied in any gauge theory.

We now show how a Lagrange multiplier (LM) field
can be used to excise all radiative effects beyond one-loop order
in this model \cite{McKeonSlavnov1992,Brandt2025a}.

\section{Using the Lagrange Multiplier}
It was shown previously that if a Lagrange multiplier
field were used to impose the classical equations of motion,
then quantum effects beyond one-loop order are eliminated
\cite{McKeonSlavnov1992}.  
However, one-loop effects present when the LM field is not present, persist.
This has made it possible to quantize the Einstein--Hilbert (EH) action
in interaction with bosonic fields in a way that is consistent with
renormalizability and unitarity \cite{Brandt:2020gms,Brandt2025a}.  
To deal with fermionic fields interacting with gravity, one must use the
Einstein--Cartan (EC) form of the metric field \cite{Kibble1961}.  
This requires special care in the quantization procedure \cite{Brandt2024a},
and so here we
we will illustrate how the LM field can be used in
conjunction with the EC action in $2+1$ dimensions of Eq.~\eqref{eq:9},
which has an additional fermionic gauge symmetry, so that this is
a type of “supergravity’’ model.

The way in which a LM field can be used to
limit radiative corrections to one-loop order can be illustrated
by considering the following integral in $n$ dimensions:
\begin{equation}
I \;=\; \int \frac{d f_{i} d\lambda_{i}}{(2\pi)^{n}}
\left|\det \frac{\partial^{2} L(f_{i})}{\partial f_{i}\partial f_{j}}\right|^{1/2}
\exp\!\left[ L(f_{i}) + i\lambda_{i}\frac{\partial L(f_{i})}{\partial f_{i}} \right] .
\label{eq:33}
\end{equation}

Upon integrating over $\lambda_{i}$, Eq.~\eqref{eq:33} becomes
\begin{equation}
I \;=\; \int d f_{i}
\left|\det \frac{\partial^{2} L(f_{i})}{\partial f_{i}\partial f_{j}}\right|^{1/2}
\exp\!\left[ L(f_{i}) \right]
\delta\!\left( \frac{\partial L(f_{i})}{\partial f_{i}} \right) .
\label{eq:34}
\end{equation}

If we now make a linear and invertible change of variables
\begin{equation}
  g_{i} = g_{i}(f_{i})
\label{eq:35}
\end{equation}
in Eq.~\eqref{eq:34}
then
\begin{equation}
d f_{i}
= \frac{d g_{i}}
{\det\!\left( \frac{\partial g_{k}}{\partial f_{l}} \right)} ,
\label{eq:36}
\end{equation}
\begin{align}
\delta\!\left( \frac{\partial L}{\partial f_{i}} \right)
&= \delta\!\left( \frac{\partial g_{j}}{\partial f_{i}}\;
                 \frac{\partial L}{\partial g_{j}} \right)
=
\frac{1}{\det\!\left( \frac{\partial g_{k}}{\partial f_{l}} \right)}\;
\delta\!\left( \frac{\partial L}{\partial g_{j}} \right) .
\label{eq:37}
\end{align}
and
\begin{equation}
\det\!\left( \frac{\partial^{2} L(f_{i})}{\partial f_{i}\partial f_{j}} \right)
=
\det\!\left[
  \left( \frac{\partial g_{k}}{\partial f_{i}} \frac{\partial }{\partial g_k}\right)
  \left( \frac{\partial g_{l}}{\partial f_{j}} \frac{\partial }{\partial g_l}\right)
  L(g_{i})
\right] .
\label{eq:38}
\end{equation}
Since
\begin{equation}
\frac{\partial L(g_{i})}{\partial g_{i}}\;
\delta\!\left( \frac{\partial L}{\partial g_{j}} \right) = 0 ,
\label{eq:40}
\end{equation}
Eqs.~\eqref{eq:36}–\eqref{eq:40} reduce Eq.~\eqref{eq:34} to
\begin{equation}
I
=
\int
\frac{d g_{i}}{\det\!\left( \frac{\partial g_{k}}{\partial f_{l}} \right)}
\left[
\det\!\left(
\frac{\partial^{2} L(g_{i})}{\partial g_{i}\partial g_{j}}
\right)
\right]^{1/2}
\delta\!\left( \frac{\partial L}{\partial g_{i}} \right)
\exp\!\big[ L(g_{i}) \big] .
\label{eq:41}
\end{equation}
We thus see that the role of the determinant factor in
Eq.~\eqref{eq:34} is to ensure that the integral $I$ is invariant
under the change of variables of Eq.~\eqref{eq:35}
\cite{Brandt2025b,Brandt2024c}.

The $\delta$–function in Eq.~\eqref{eq:34} makes it possible to explicitly
perform the integral over $f_{i}$  leading to
\begin{equation}
I
= \sum_{\alpha}
\left[
\det\!\left(
\frac{\partial^{2} L(f_{i}^{\alpha})}{\partial f_{i}\partial f_{j}}
\right)
\right]^{-1/2}
\exp\!\left[ L(f_{i}^{\alpha}) \right] ,
\label{eq:42}
\end{equation}
where $f_{i}^{\alpha}$ is a solution to the equation
\begin{equation}
\frac{\partial L(f_{i}^{\alpha})}{\partial f_{i}} = 0 .
\label{eq:43}
\end{equation}

One can generalize this sequence of steps so that
the integral in Eq.~\eqref{eq:33} becomes a path integral, with the
variables $f_{i}$ becoming fields, the parameters $\lambda_{i}$ a LM field,
and $L(f_{i})$ the classical action.  
In Eq.~\eqref{eq:42}, the exponential
$\exp[L(f_{i}^{\alpha})]$ becomes the sum of all tree–level diagrams in the
presence of solutions of the classical equation of motion $f_{i}^{\alpha}$
of Eq.~\eqref{eq:43} \cite{Brandt2024c}, and the functional determinant
$\det^{-1/2}\!\left( \frac{\partial^{2} L}{\partial f_{i}\partial f_{j}} \right)$
is the contribution of all one–loop
diagrams in the presence of $f_{i}^{\alpha}$.  

We are able to treat the path integral
\begin{equation}\begin{split}
&Z[ j_{i}(x), k_{i}(x), J_{I}(x) ]
=\int  \mathcal{D}f_{i}(x) \mathcal{D}\lambda_{i}(x) \mathcal{D}\Phi_{I}(x)
 \left|\det\!\left( \frac{\partial^{2} {\cal L}_f(f_{i})}{\partial f_{i}\partial f_{j}} \right)\right|^{1/2}
\\ & \times \exp\!\left\{ i \!\int d x\left[
{\cal L}_f(f_{i})
+ \lambda_{i}\frac{\partial {\cal L}_f(f_{i})}{\partial f_{i}}
+ {\cal L}_{\Phi}(\Phi_{i},f_{i})
+ j_{i}f_{i}
+ k_{i}\lambda_{i}
+ J_{I}\phi_{I}
\right]\right\}
\end{split}
\label{eq:44}
\end{equation}
using the functional analogues of the steps used to obtain
Eq.~\eqref{eq:42}, to show that  
\begin{equation}
\begin{split}
Z[ j_{i}(x), k_{i}(x), J_{I}(x) ]&
= \int \mathcal{D}\Phi_{I}(x)
\sum_{a}
\left|\det\!\left( \frac{\partial^{2}{\cal L}_f(f_{i}^{a})}{\partial f_{i}\partial f_{j}} \right)\right|^{-1/2}\\
& \times
\exp\!\left[i\!\int d {x}{\cal L}_f(f_{i}^{a})\right]
\exp\!\left[i\!\int d {x}{\cal L}_{I}(\Phi_{I},f_{i}^{a})\right]
\exp\!\left[i\!\int d {x}\big(j_{i}f_{i}^{a} + J_{I}\Phi_{I}\big)\right],
\end{split}
\label{eq:45}
\end{equation}
where $f_{i}^{a}(x)$ satisfy the equation
\begin{equation}
\frac{\partial {\cal L}_f(f_{i}^{a})}{\partial f_{i}} + k_{i} = 0 .
\label{eq:46}
\end{equation}
Schematically, the fields $f_{i}$ can be identified with the graviton field
and $\Phi_{I}(x)$ with a bosonic matter field, with ${\cal L}_{f}$ being the
EH Lagrangian.  
We see from Eq.~\eqref{eq:45} that all loop effects
involving virtual graviton field contributions are confined to
one–loop order, being given by the functional determinant that
occurs explicitly.  
Divergences that arise all vanish on shell \cite{tHooftVeltman1974},
and so $Z$ is free of divergences arising from the propagation of the
gravitational field.  
Divergences arising from propagation of the matter field
$\Phi_{I}$ can be absorbed using conventional renormalization
\cite{WeinbergQTF1996}.  
The contribution of $f_{i}^{a}$ is much like that of “dark matter,”
as it is a purely gravitational contribution to the dynamics of the
matter field.

In place of this derivation of $Z$, one could use a
conventional perturbative approach to evaluating Eq.~\eqref{eq:44}.
This
involves introducing background fields \cite{Abbott1981},
\begin{equation}
\bar{f}_{i} = \frac{\partial W(j_{i},k_{i},J_{I})}{\partial j_{i}},
\qquad
\bar{\lambda}_{i} = \frac{\partial W(j_{i},k_{i},J_{I})}{\partial k_{i}},
\qquad
\bar{\Phi}_{i} = \frac{\partial W(j_{i},k_{i},J_{I})}{\partial J_{I}} .
\label{eq:47}
\end{equation}
where
\begin{equation}
W(j_{i},k_{i},J_{I})
= -i\ln Z(j_{i},k_{i},J_{I}).
\label{eq:48}
\end{equation}
Upon making the Legendre transform
\begin{equation}
\Gamma(\bar{f}_{i},\bar{\lambda}_{i},\bar{\Phi}_{I})
= W(j_{i},k_{i},J_{I})
- \int d{x}\big(
j_{i}\bar{f}_{i}
+ k_{i}\bar{\lambda}_{i}
+ J_{I}\bar{\Phi}_{I}
\big),
\label{eq:49}
\end{equation}
we see that Eq.~\eqref{eq:44} becomes
\begin{align}
&\exp\!\left[ i\Gamma(\bar{f}_{i},\bar{\lambda}_{i},\bar{\Phi}_{I}) \right]
=
\int \mathcal{D}q_{i}\mathcal{D}\sigma_{i}\mathcal{D}Q_I\;
\left|\det\!\left(
\frac{\partial^{2}{\cal L}(\bar{f}_{i}+q_{i})}{\partial q_{i}\partial q_{j}}
\right)\right|^{1/2}
\label{eq:50}
\\
&times
\exp\!\left\{ i\!\int d{x}
\left[
{\cal L}_{f}(\bar{f}_{i}+q_{i})
+ (\bar{\lambda}_{i}+\sigma_{i})
\frac{\partial {\cal L}_{f}(\bar{f}_{i}+q_{i})}{\partial g_{i}}
\right.\right.
\nonumber \left.\left.
+{\cal L}_{I}(\bar{\Phi}_{I}+Q_{I},\bar{f}_{i}+q_{i})
+ j_{i}q_{i}
+ k_{i}\sigma_{i}
+ J_{I}Q_{I}
\right]\right\}.
\nonumber
\end{align}
Integration over $\sigma_{i}$ yields a functional $\delta$–function,
and this in turn allows one to integrate over $q_{i}$, leading to
\begin{align}
&\exp\!\left[ i\Gamma(\bar f_i,\bar\lambda_i,\bar\Phi_I) \right]
=
\int \mathcal{D}Q_I\;
\left|\det\!\left(
\frac{\partial^{2} {\cal L}_f(\bar f_i + q^a_i)}{\partial q_i\partial q_j}
\right)\right|^{-1/2}
\label{eq:51}
\\
&\qquad\times
\exp\!\left\{ i\!\int d x
\left[
{\cal L}_f(\bar f_i + q^a_i)
+ \bar\lambda_i\frac{\partial {\cal L}_f(\bar f_i + q^a_i)}{\partial g_i}
\right.\right.
\left.\left.
+ {\cal L}_\Phi(\bar\Phi_I + Q_I,\;\bar f_i + q^a_i)
+ j_i \bar{q}^a_i
+ J_I Q_I
\right]\right\}.
\nonumber
\end{align}
where \(q_i^{a}\) satisfies the classical equation of motion
\begin{equation}
\frac{\partial {\cal L}_f(\bar f_i + q_i^{a})}{\partial g_i} = 0 .
\label{eq:52}
\end{equation}
when \(k_i = 0\).  
Since when dealing with EH theory where
there are bosonic matter fields, divergences arising from the
propagation of gravitons vanish when Eq.~\eqref{eq:52} is satisfied
\cite{tHooftVeltman1974}, then
these divergences can be absorbed by renormalizing \(\bar{\lambda}_i\).
Unitarity can also be established \cite{Brandt2024b}.

In this outline of how the LM field can be used, we have ignored the
difficulties that arise when there is a gauge symmetry. Let us
generalize so that there are $M$ fields $f_i^{A}$ $(A=1,\ldots,M)$ and
$N$ gauge functions $\xi^{\Gamma}_i$ $(\Gamma=1,\ldots,N)$, and set the
background fields $\bar{\Phi}_I$ equal to zero $(\bar{\Phi}_I = 0)$.

Invariance under the gauge transformation
\begin{equation}
f_i^{A} \;\to\; f_i^{A} + R^{A\Gamma}_{ij}(f^A_i)\xi^{\Gamma}_j,
\label{eq:5.21}
\end{equation}
means that
\begin{equation}
\mathcal{L}_f(f_i^{A})
= \mathcal{L}_f\!\left(f_i^{A} + R^{A\Gamma}_{ij}\xi^{\Gamma}_j\right),
\label{eq:5.22}
\end{equation}
so that
\begin{equation}
\frac{\partial \mathcal{L}_f}{\partial f_i^{A}}
R^{A\Gamma}_{ij} \xi^\Gamma_j = 0 .
\label{eq:5.23}
\end{equation}
Consequently, 
$
\mathcal{L}_f
+ \lambda_i^{A}
\dfrac{\partial \mathcal{L}_f}{\partial f_i^{A}}
$
is invariant if
\begin{equation}
\lambda_i^{A} \;\to\;
\lambda_i^{A} + R^{A\Gamma}_{ij}(f^A_i)\zeta^{\Gamma}_j,
\label{eq:5.24}
\end{equation}
for a gauge function $\zeta^{\Gamma}_j$. So long as
$\mathcal{L}_f$ satisfies Eq.~\eqref{eq:5.22}, then
$
\mathcal{L}_f(f_i^{A})
+ \lambda_i^{A}
\frac{\partial \mathcal{L}_f(f_i^{A})}{\partial f_i^{A}}
$
is invariant if $f_i^{A}$ transform as in
Eq.~\eqref{eq:5.21}, and $\lambda_i^{A}$ transforms according to \cite{Brandt:2020gms}
\begin{equation}
\lambda_i^{A'} =
\lambda_j^{B}
\frac{\partial f_i^{A'}}{\partial f_j^{B}},
\end{equation}
which by Eq.~\eqref{eq:5.21} becomes
\begin{equation}
\lambda_i^{A} \;\to\;
\lambda_i^{A}
+ \lambda_l^{B}
\frac{\partial R^{A\Gamma}_{ij}}{\partial f_l^{B}}
\xi^{\Gamma}_j.
\label{eq:5.25}
\end{equation}


The Faddeev–Popov (FP) procedure can be used
to eliminate the redundancies in the path integral of Eq. \eqref{eq:50}
that occur because of these gauge invariances \cite{FaddeevPopov1967}. (See also
Ref. \cite{Nielsen1975}.) Normally one selects a “gauge condition’’ to be
imposed on the field $f_i$, but in fact one may impose more than
one such condition (such as having the graviton field being both
transverse and traceless \cite{Brandt2024d}). This involves the $n$–conditions
\begin{equation}
A^{(\alpha)}_{ij} f_i = 0 = A^{(\alpha)}_{i j} \lambda_j \qquad (\alpha = 1,2,\ldots ,n)
\label{eq:58}
\end{equation}
and inserting a constant factor of
\begin{equation}
\begin{aligned}
&\int {\cal D}\xi^{(\alpha) A}_{i} {\cal D}\zeta^{(\alpha) A}_{i}\;
\delta\!\Bigg[
A^{(\alpha)}_{ij}
\begin{pmatrix}
f_{j} \\[3pt]
\lambda_{j}
\end{pmatrix}
+
\begin{pmatrix}
0 & R^{AB}_{ij} \\[4pt]
R^{AB}_{ij} &
\lambda_l\dfrac{\partial R^{AB}_{ij}}{\partial f_{l}}
\end{pmatrix}
\begin{pmatrix}
\zeta^{(\alpha) B}_{j} \\[4pt]
\xi^{(\alpha) B}_{j}
\end{pmatrix}
-\;
\begin{pmatrix}
p^{(\alpha)}_{i} \\[4pt]
q^{(\alpha)}_{i}
\end{pmatrix}
\;
\Bigg]
\det A^{(\alpha)}_{ij}
\begin{pmatrix}
0 & R^{AB}_{jk} \\[4pt]
R^{AB}_{jk} &
\lambda_l\dfrac{\partial R^{AB}_{jk}}{\partial f_{l}}
\end{pmatrix}
\end{aligned}
\label{eq:59}
\end{equation}
into the path integral of Eq. \eqref{eq:44}.

We now will consider only a single gauge condition
(i.e., take $n=1$ in eq. \eqref{eq:58}) and work with the model of
Eq. \eqref{eq:9}. This entails having to consider not just bosonic
(commuting) fields $b_{\mu a}$ and $\omega_{\mu a}$ but also a fermionic
(anticommuting or Grassmann) field $\psi_{\mu}$.  There will also be
a pair of bosonic LM fields $l_{\mu a}$ and $L_{\mu a}$, as well as a
fermionic LM field $\lambda_{\mu}$.  The three gauge transformations are
those of Eq. \eqref{eq:10}.  We make the replacement
\begin{align}
\nonumber\label{eq:60}
{\cal L}_{f}(f_{i}^{A}) + \lambda_{i}\frac{\partial {\cal L}_{f}(f_{i}^{A})}{\partial f_{i}^{A}}
\;\longrightarrow\;&
{\cal L}_{EC3}(b_{\mu i},\omega_{\mu i},\psi_{\mu})
+ l_{\mu i}\frac{\partial {\cal L}_{EC3}}{\partial b_{\mu i}}
+ L_{\mu i}\frac{\partial {\cal L}_{EC3}}{\partial \omega_{\mu i}}
+ \bar{\lambda}_{\mu}\frac{\partial {\cal L}_{EC3}}{\partial \bar{\psi}_{\mu}}\\
\nonumber
&= \varepsilon^{\mu\nu\lambda}
\left( b_{\mu}^{i} R_{\nu\lambda i}
+ \bar{\psi}_{\mu} D_{\nu}\psi_{\lambda}
+ \frac{\Lambda}{3}\varepsilon_{ijk} b_{\mu}^{i} b_{\nu}^{j} b_{\lambda}^{k}
+ \frac{i\varkappa}{2} \bar{\psi}_{\mu} b_{\nu}^{i}\gamma_{i}\psi_{\lambda} \right)
\\\nonumber
&
+ l^{\mu i}\varepsilon_{\mu\nu\lambda}
  \left(R^{\nu\lambda}_{\;\;\; i} 
  + {\Lambda}\varepsilon_{ijk} b^{\nu j}  b^{\lambda k} 
  - \frac{i\varkappa}{2} \bar{\psi}^{\nu} \gamma_{i}\psi^{\lambda} \right)
\\
&
+ 2 L^{\mu i}\varepsilon_{\mu\nu\lambda}
  \left( \varepsilon_{ijk}{\cal D}^{\nu j} b^{\lambda k} -\frac{i}{4} \bar{\psi}^\nu \gamma_i \psi^\lambda\right)
  + 2 \bar{\lambda}^{\mu}\varepsilon_{\mu\nu\lambda}
  \left( {D}^{\nu} \psi^{\lambda} +\frac{i\varkappa}{2} b^{\nu i} \gamma_i \psi^\lambda\right)
\end{align}
with Eq. \eqref{eq:11} being satisfied.
Using Eqs. \eqref{eq:5.24} and \eqref{eq:5.25}, we can see that the gauge
invariances of Eq. \eqref{eq:60} result in
\begin{subequations}
\begin{align}
\delta l_{\mu}^{i} &= {\cal D}_{\mu}^{i j}\tilde{A}_{j}
     - \varepsilon^{i j k} b_{\mu j}\tilde{B}_{k}
     + \frac{i}{2}\tilde{\bar{C}}\gamma^{i}\psi_{\mu}
 -\varepsilon^{i j k} l_{\mu j} B_{k}
 - \varepsilon^{i j k} L_{\mu j} A_{k}
 - \frac{i}{2}\bar{\lambda}_{\mu}\gamma^{i} C, \\
\delta L_{\mu}^{i} &=\varepsilon^{ijk} {\cal D}_{\mu j}\tilde{B}_{k}
   - \varkappa^{2} \varepsilon^{i j k} b_{\mu j}\tilde{A}_{k}
   + \frac{i\varkappa}{2}\tilde{\bar{C}}\gamma^{i}\psi_{\mu}
 - \varkappa^{2}\varepsilon^{i j k} l_{\mu j} B_{k}
 - \varepsilon^{i j k} L_{\mu j} B_{k}
 - \frac{i\varkappa}{2}\bar{\lambda}_{\mu}\gamma^{j} C,\\
\delta \lambda_\mu &= D_{\mu} \tilde{C}
   + \frac{i\varkappa}{2} \gamma^j b_{\mu j}\tilde{C}
   - \frac{i \gamma^j}{2}(\varkappa \tilde{A}_{j} + \tilde{B}_{j} ) \psi_{\mu}
 + \frac{i\varkappa}{2} \gamma^j l_{\mu j} C
 + \frac{i \gamma^j}{2} L_{\mu j} C
 - \frac{i\gamma^{j}}{2}(\varkappa A_{ j} + B_{j})\lambda_\mu ;
\end{align}
\end{subequations}
where \((\tilde{A}_{i}, \tilde{B}_{i}, \tilde{C}, A_{i}, B_{i}, C)\) are gauge functions.

A straightforward but tedious application of the
formalism for quantizing gauge systems with LM fields
can now be made, following the formalism outlined in \cite{Brandt2024c}.
The only additional consideration is to accommodate more than
one gauge invariance in the classical action, such as occurs in
Eq. \eqref{eq:10}.

\section{Discussion}

We have outlined several aspects of supergravity in $2+1$ dimensions, focusing
in particular on the structure of its gauge symmetries, the quantization of the
model using the Faddeev--Popov--Nielsen procedure, and the role of Lagrange
multiplier fields in eliminating quantum effects beyond one-loop order.
The $2+1$ Einstein--Cartan model provides a setting in which these
features can be analyzed without resorting to background fields, and in which
the full set of first-class constraints and their associated gauge
transformations can be determined explicitly.

When extending these results to $3+1$ dimensions, a central open
problem lies in identifying the complete gauge structure
that follows from the canonical formulation of the four-dimensional EC theory,
and understanding how these symmetries can be consistently implemented in the
quantized theory \cite{Brandt2024a}. Once this is achieved, the Lagrange
multiplier mechanism should again make it possible to construct a model in
which the metric field, when coupled to both bosonic or fermionic matter,
remains both renormalizable and unitary. This has already been demonstrated in
the case in which the metric interacts solely with bosonic fields
in $3+1$D
\cite{Brandt2025a}, using the Einstein--Hilbert action.

A formulation in which the metric couples consistently to both species of
matter fields would provide a natural pathway toward extending the Standard
Model so as to incorporate gravitational interactions on the same footing as
the strong and electroweak interactions. A proposal for such a unified
framework has recently been advanced in Ref.~\cite{Chishtie2025}. The analysis
presented here offers a concrete lower-dimensional template for this program
and illustrates how the Lagrange multiplier formalism can coexist with
nontrivial gauge structures while restricting radiative effects to one-loop
order.

\begin{acknowledgments}
 F.\ T.\ B., J.\ F.\ thank CNPq (Brazil) for financial support.  This study was financed, in part, by the São Paulo Research Foundation (FAPESP), Brasil. Process Number \#2025/16156-7. S.\ M.-F. thanks CNPq for partial financial support.
\end{acknowledgments}

\appendix

\section{Notation}\label{appA}
In Eq.~\eqref{eq:9} we use the metric 
$
\eta_{ij} = \mathrm{diag}(+ - -),
$
and the Dirac matrices $(\gamma^{0},\gamma^1,\gamma^2) = (\sigma_{2}, i \sigma_{1},i \sigma_{3})$.
They satisfy
\begin{equation}
\gamma^{i}\gamma^{j} 
=  \eta^{ij} + i \varepsilon^{ijk} \gamma_k; \;\;\; (\varepsilon^{012} = +1)
\label{eq:A1}
\end{equation}
Grassmann spinors are Majorana so that
\begin{equation}
\psi = -\gamma^{0} \bar{\psi}^{T} = \psi^{\ast},
\qquad
(\bar{\psi} = \psi^{\dagger} \gamma^{0}),
\label{eq:A2}
\end{equation}
with
\begin{equation}
\bar{\psi}\chi = \bar{\chi}\psi,
\qquad
\bar{\psi}\gamma^{i}\chi = -\bar{\chi}\gamma^{i}\psi.
\label{eq:A3}
\end{equation}

We also have
\begin{subequations}
\begin{equation}
D_{\mu} = \partial_{\mu} + \frac{i}{2} \gamma^i \omega_{\mu i},
\label{eq:A4a}
\end{equation}
\begin{equation}
{\cal D}_{\mu}^{ij} 
= \partial_{\mu} \delta^{ij}  
- \varepsilon^{imj} \omega_{\mu m},
\label{eq:A4b}
\end{equation}
\begin{equation}
R_{\mu\nu i} 
= \partial_{\mu}\omega_{\nu i}
- \partial_{\nu}\omega_{\mu i}
- \varepsilon_{ijk} \omega_{\mu}^{ j}\omega_{\nu}^{ k},
\label{eq:A4c}
\end{equation}
\end{subequations}
as well as the usual connection between the
``dreibein'' $e_{\mu i}$ and metric $g_{\mu\nu}$:
\begin{equation}
g_{\mu\nu} = \eta_{ij} e_{\mu}^{\ i} e_{\nu}^{\ j},
\qquad
\eta_{ij} = g^{\mu\nu} e_{\mu i} e_{\nu j} \; ,
\label{eq:A5}
\end{equation}
where $e_{\mu a} = L b_{\mu a}$ with the dimensionfull parameter $L$ being scaled to equal one. The dimensionless parameter $\varkappa$ is related to $L$ according to $ \varkappa = \kappa/L$, where $\kappa^2 = 16 \pi G$.

\end{document}